# New Result for the $p^7\text{Be}\rightarrow{}^8\text{B}\gamma$ Astrophysical *S*-factor from 10 keV to 5 MeV and Reaction Rate from 0.01 to 10 $T_9$

Dubovichenko S.B.[1,2,*], Burkova N.A.[2], Dzhazairov-Kakhramanov A.V.[1,*], Tkachenko A.S.[1,2]

[1] Fesenkov Astrophysical Institute "NCSRT" ASA MDASI RK, 050020, Almaty, RK
[2] al-Farabi Kazakh National University, RK, 050040, Almaty, RK
E-mail: dubovichenko@mail.ru, albert-j@yandex.ru

**Abstract**

The astrophysical *S*-factor for the process $p^7\text{Be} \rightarrow {}^8\text{B}\gamma$ of radiative capture by the $^8$B ground state is described up to 5 MeV, within the modified potential cluster model. The signatures of the *S*-factor resonances at 0.632, 2.18 and 3.36 MeV due to *M*1 and *E*2 transitions from the scattering of resonance $^3P_1$, $^3F_3$ and $^3D_2$ waves to the ground $^3P_2$ state are illuminated using total cross sections, astrophysical *S*-factors and reaction rates at temperatures from 0.01 to 10.0 $T_9$. Experimental data on the *S*-factor are reproduced well in the range 100 keV to 3 MeV. The importance of the input of the 0.632 MeV resonance to the reaction rate is proved. Our calculations (performed in advance) confirm the latest data for the *S*-factor at 19 keV.

## 1. Introduction

The radiative capture process $p+{}^7\text{Be}\rightarrow{}^8\text{B}+\gamma$ is one of the final reactions in the *pp*-chain [1], providing energy for the Sun's activity and similarly for the majority of stars in the Universe. The unstable product nucleus $^8$B weakly decays via $^8\text{Be}+e^++\nu$ forming $^8$Be, and ends with two α-particles, $^4\text{He} + {}^4\text{He}$. The high-energy neutrinos from $^8$B decay are detected on Earth, providing a direct reflection of this process occurring in the Sun [2].

The modified potential cluster model (MPCM) [3,4] is used to treat the proton capture reaction involving $^7$Be within the energy range 10 keV to 5 MeV, in addition to its application in the solution of similar astrophysical problems [4,5]. A classification based on orbital symmetry was used in the construction of the $p^7$Be interaction potentials responsible for correct reproduction of the discrete and continuous spectra, and has also been used successfully for other clustered systems [6–9].

The continuum potential parameters are fitted to the experimental phase shifts, extracted from the fit of the scattering differential cross sections or the excitation spectra of the final nuclei. The study published in this paper was previously published in a brief form in [10], although in that manuscript only certain parts of the *E*1 and *M*1 transitions were calculated and very narrow energy range was considered.

Both the ground and excited bound states were fixed at a high quantitative level, reproducing the available observed parameters in cluster reaction channels [11,12]. Some cases have been reconfirmed within three-body systems, as in [13].

---
[*] Corresponding authors

$^8$B is a stable nucleus in terms of strong interactions, as it disintegrates in a transition to $^8$Be only at the expense of weak interactions. It is therefore quite possible to consider $^8$B as a $p^7$Be cluster system and to apply the known methods of the MPCM to it [4,5]. Consequently, one of goals of this work is to explore the possibilities of the MPCM and to adapt it to explain the basic characteristics of the $p^7$Be system and the radiative proton capture by $^7$Be at astrophysical energies.

The second goal pursued in this work is to clarify the role of the first maximum in the total cross sections, or astrophysical *S*-factor, and to determine its contribution to the reaction rate. Consequently, the significant contribution of the first resonance at 0.632 MeV to the reaction rate of the radiative proton capture by $^7$Be at specific temperatures is demonstrated here.

## 1.1. Astrophysical aspects

In the study of many astrophysical objects, such as the primordial Universe, stellar evolution, supernovae etc., an important input parameter in the astrophysical models used is the total cross section of the radiative capture reactions. Since the temperature is relatively small in many of these media and systems, these cross sections must be obtained at very low energies, which range from tens to hundreds of keV. Only a few of these sections can be measured in the laboratory; most are obtained from extrapolation to lower energies. In particular, for direct proton capture involving radioactive nuclei such as $^8$B or $^7$Be, direct measurement of the cross sections is very difficult, due to their small size and the limited beam intensity of these radioactive elements.

Indirect methods of measurements have been developed for reactions of a similar type, for example the Coulomb dissociation method (see e.g. [14]). However, similar, indirect methods of measurement lead to large experimental errors. This in turn leads to large ambiguities in results obtained on the basis of using astrophysical models to describe different processes in the Universe. In particular, extrapolation of the *S*-factor to low energies is not a direct method, and the errors in this method are unpredictable [15].

In view of these experimental limitations, theoretical models of nuclear reactions are obviously a necessary addition to experimental studies. Theoretical calculations can be performed at any energy, and are not limited by the instability of the nuclei. However, they also face many other problems; for example, the radiative capture reactions associated with stellar nucleosynthesis are often difficult to analyze. It is impossible for any nuclear model to represent the thousands of reactions associated with modern models of the evolution of stars. Nevertheless, a limited number of reactions, particularly in the low-mass range, play an important role and deserve special attention [16].

These include the reaction considered in this work, since it is closely related to the recording of solar neutrinos on Earth. Its detailed study, and in particular the form and dependence on energy of the astrophysical *S*-factor, the reaction rate, and its dependence on low-lying resonances, are therefore of considerable interest in many problems of nuclear astrophysics. The results will contribute to a better general understanding of the mechanisms of thermonuclear processes occurring in various synthesis reactions.

## 1.2. Physical aspects

Since the reaction $p^7Be \rightarrow {}^8B\gamma$ is important both for astrophysical applications and for a general understanding of the physics of thermonuclear processes, experimental and theoretical studies on this topic are ongoing. There have been so many works in this area that it would be impossible to review them in a regular article, and difficult even in a long review. We therefore focus here only on some recent studies of this reaction.

Recent experimental data [17] have been reported from measurements of the total cross sections and astrophysical *S*-factor in the energy range 0.12–1.75 MeV in the center of mass (c.m.) system. A resulting value for the *S*-factor at zero energy of $S_{17}(0) = 21.5 \pm 0.6(\exp) \pm 0.7(\text{theor})$ eV·b was obtained. After comparison with other results, an average value of $S_{17}(0) = 20.9 \pm 0.6(\exp) \pm 0.7(\text{theor})$ eV·b was suggested.

In [18], the reaction rates at temperatures from 0.003 to 10 $T_9$ were calculated based on a highly accurate approximation of the *S*-factor. A value of 0.57(10) fm$^{-1}$ was obtained for the asymptotic normalization coefficient (ANC), and for the astrophysical *S*-factor at zero energy, a value 19.9(3.5) eV·b was found, which is within the error limits of most of the previous results, including the data in [17]. It was shown that the first resonance in the p$^7$Be system at 0.632 MeV in c.m. makes a significant contribution to the reaction rate at a temperature of about 2 $T_9$.

Comparatively recently, an analysis of the experimental data for the $^7Be(d,n)^8B$ reaction was carried out in [19], and the value of the ANC p$^7$Be channel was found to be 0.613(60) fm$^{-1}$, which corresponds to an *S*-factor of 22.8(2.2) eV·b at zero energy. A detailed table of results is given that compares the results with those of a dozen other works on the ANC and *S*-factor obtained over the last two decades.

An astrophysical *S*-factor of 19.4(7) eV·b for zero energy was obtained in [20] on the basis of an *ab initio* approach. This value has a minimal error compared with previous calculations, and agrees well with the average results of approximations of experimental data from [17]. However, the resonance at 0.632 MeV (c.m.) was not considered in [20] or in previous analogous work [21]. The authors of these studies believed that this resonance would not make a significant contribution to the reaction rates, although this is strongly refuted in [18].

In [22], the p$^7$Be capture reaction is considered within the framework of effective field theory. A value of 21.3(7) eV·b was obtained for the *S*-factor at zero energy, leading to an uncertainty half as large as that previously given in earlier studies. The proposed value, within the limits of errors, coincides with the results obtained in previous *ab initio* calculations [20]. In addition, it agrees with the average value for the *S*-factor proposed in [17], on the basis of a review of various experimental data obtained in different papers over the last 20–30 years.

Although the total cross sections of such processes can be calculated on the basis of various theoretical models, the accuracy of such calculations is not always clear (see e.g. [23]). It is therefore important to choose a nuclear model that can describe the main characteristics of reactions, for example radiative capture based on certain unified principles, criteria and methods. The nuclear model used in the present work has already made it possible to describe the main characteristics of more than 30 radiative capture reactions [4–13,24]. It is therefore likely that it will give an acceptable description of the reaction considered here.

## 2. Model concept
## 2.1. Principles of building potentials

Earlier, within the framework of a modified potential cluster model with forbidden states (FS), we demonstrated the possibility of describing the astrophysical *S*-factors of radiative capture reactions for many light (and lightest) atomic nuclei [4,5,25]. This model takes into account the super-multiplet symmetry of the cluster wave function with the separation of the orbital states, according to Young diagrams.

This classification allows us to analyze the structure of the intercluster interactions, and to determine the number of allowed states (AS) and forbidden states in relevant wave functions, thereby making it possible to find the number of nodes of the radial wave function of the relative cluster motion. For any cluster system, the multiparticle nature of the problem and the effects of antisymmetrization are qualitatively taken into account by the separation of the single-particle bound levels of the corresponding potential into states that are allowed or forbidden by the Pauli principle [3].

The present approach allows us to construct partial interaction potentials for the scattering processes, depending on the quantum numbers *L, S, J,* and in some cases on Young diagrams {*f*}, on the basis of a description of the elastic scattering phase shifts and taking into account their resonance behavior.

These phase shifts are extracted from the experimental differential cross sections using phase shift analysis. For the bound states of light nuclei in cluster channels, the potentials are constructed not only on the basis of the description of the scattering phase shifts, but also using certain additional requirements. For example, one such requirement is the reproduction of the binding energy and other basic characteristics of the bound states of the nuclei. It is assumed that the bound state of the nucleus is mainly due to a cluster channel consisting of initial particles that participate in the capture reaction [6–8].

As a result, not all of the potentials contain the ambiguities inherent in the optical model [26], nor allow the correct description of the total cross sections of radiative capture processes for $p^7$Be system, as will be seen below. The potentials of the bound states should correctly describe the known values of asymptotic constant (AC), *C*. This is related to the asymptotic normalization coefficient (ANC), which is usually derived from experiment and which we denote by $A_{NC}$, as follows [27]:

$$A_{NC}^2 = S_f \cdot C^2, \qquad (1)$$

where $S_f$ is the spectroscopic factor, and *C* is the dimension AC in fm$^{-1/2}$, defined in the relation

$$\chi_L(r) = C W_{-\eta L+1/2}(2k_0 r). \qquad (2)$$

Here, $\chi_L(r)$ is the numerical GS radial wave function, obtained as the solution of the Schrödinger equation normalized to unit size, $W_{-\eta L+1/2}(2k_0 r)$ is the Whittaker function, $k_0$ is a wave number related to the channel binding energy, $\eta$ is the Coulomb parameter $\eta = \dfrac{\mu Z_1 Z_2 e^2}{\hbar^2 k_0} = 3.44476 \cdot 10^{-2} \dfrac{\mu Z_1 Z_2}{k_0}$, $Z_1$ and $Z_2$ are the particle charges, and $\mu = m_1m_2/(m_1+m_2)$ is the reduced mass of two particles and $k_0^2 = \dfrac{2\mu E}{\hbar^2}$.

Similarly, the dimensionless AC $C_w$ is defined by [28]:

$$\chi_L(r) = \sqrt{2k_0}\, C_w W_{-\eta L+1/2}(2k_0 r). \tag{3}$$

Hence, an obvious relation between the constants used here is $C = \sqrt{2k_0}\, C_W$.

## 2.2. Calculation method for the total cross sections

The total cross sections $\sigma(NJ, J_f)$ of the radiative capture via the $EJ$ and $M1$ transitions within the potential cluster model are given, for example, in [4,5] and [29] as follows:

$$\sigma_c(NJ, J_f) = \dfrac{8\pi K e^2}{\hbar^2 q^3} \dfrac{\mu \cdot A_J^2(NJ, K)}{(2S_1+1)(2S_2+1)} \dfrac{J+1}{J[(2J+1)!!]^2} \sum_{L_i, J_i} P_J^2(NJ, J_f, J_i) I_J^2(J_f, J_i), \tag{4}$$

where $\sigma$ is the total cross section of the radiative capture process, $\mu$ is the reduced mass of particles in the entry channel, $q$ is the wave number in entry channel, $S_1$ and $S_2$ are the spins of particles in the initial channel, $K$ and $J$ are the wave number and angular momentum of the $\gamma$-quantum, and $N$ is the electrical ($E$) or magnetic ($M$) transitions of the multipolarity $J$ from the initial continuous state $J_i$ to the final nuclear state $J_f$.

For the electrical convection $EJ(L)$ transitions $(S_i = S_f = S)$, the following expressions hold for $P_J$, $A_J$ and $I_J$ [4,5]:

$$P_J^2(EJ, J_f, J_i) = \delta_{S_i S_f}[(2J+1)(2L_i+1)(2J_i+1)(2J_f+1)](L_i 0 J 0 | L_f 0)^2 \begin{Bmatrix} L_i & S & J_i \\ J_f & J & L_f \end{Bmatrix}^2$$

$$A_J(EJ, K) = K^J \mu^J \left(\dfrac{Z_1}{m_1^J} + (-1)^J \dfrac{Z_2}{m_2^J}\right), \quad I_J(J_f, J_i) = \langle \chi_f | r^J | \chi_i \rangle. \tag{5}$$

Here, $S_i$, $S_f$, $L_f$, $L_i$, $J_f$ and $J_i$ are the corresponding momenta in the initial ($i$) and final channels ($f$), $m_1$ and $m_2$ are the particle masses in amu, $Z_1$ and $Z_2$ are the charges in units of elementary charge $e$, and $I_J$ is the overlapping integral over the radial functions for the scattering $\chi_i$ and bound $\chi_f$ states, depending on the cluster-cluster relative coordinate $r$.

For the spin dipole magnetic $M1(S)$ transition, i.e. for $J = 1$, the following expressions were obtained $(S_i = S_f = S, L_i = L_f = L)$ [4,5]:

$$P_1^2(M1, J_f, J_i) = \delta_{S_i S_f}\delta_{L_i L_f}[S(S+1)(2S+1)(2J_i+1)(2J_f+1)]\begin{Bmatrix} S & L & J_i \\ J_f & 1 & S \end{Bmatrix}^2 =$$

$$= \delta_{S_i S_f}\delta_{L_i L_f}[S(S+1)(2S+1)(2J_i+1)(2J_f+1)]\begin{Bmatrix} L & S & J_i \\ 1 & J_f & S \end{Bmatrix}^2,$$

$$A_1(M1,K) = i\frac{\hbar K}{m_0 c}\sqrt{3}\left[\mu_1\frac{m_2}{m} - \mu_2\frac{m_1}{m}\right], \qquad I_J(J_f,J_i) = \langle\chi_f|r^{J-1}|\chi_i\rangle. \qquad (6)$$

Here $m$ is the mass of the nucleus in amu, $\mu_1$, $\mu_2$ are the magnetic moments of the clusters, and other notation is the same as in the previous expression.

In all our calculations, the exact values of the particle masses $m(^7\text{Be}) = 7.014735$ amu [30] and $m_p = 1.0072764669$ amu [31] were specified. The constant $\hbar^2/m_0$ is assumed to be 41.4686 MeV·fm$^2$ and the Coulomb potential is chosen in the point-like form $V_c(\text{MeV})=1.439975\cdot Z_1\cdot Z_2/r$, where $r$ is the distance between the particles in fm. We give these values since, in addition to the masses of the nuclei, these substantially affect the binding energy of the particles in $^8$B and the resonance energy in the scattering processes. For the magnetic moments of the a proton and $^7$Be, we use the values $\mu_p = 2.792847$ and $\mu(^7\text{Be}) = -1.398(15)$ from [32].

## 2.3. Classification of cluster states in the N$^7$Be system

Firstly, we remark that the isospin $T = 1$ in the $p^7$Be system due to $T_z = +1$ [33], so that this channel is pure by $T$ unlike the $p^7$Li system [6], and $T_z = -1$ and $T = 1$ for the cluster system $n^7$Li, forming the isobaric triplet. This dictates the further classification of $p^7$Be system in bound states using orbital Young diagrams.

The allowed orbital diagram $\{f\}_L$ of a nucleus $A\{f\}$ associated as two sub-systems $A_1(\{f_1\})+A_2(\{f_2\})$ is the direct external product $\{f\}_L = \{f_1\}_L \times \{f_2\}_L$ defined by the Littlewood theorem [34]. The antisymmetrization conditions for the total wave function of the $A = A_1 + A_2$ system [3] are the allowed or forbidden states according to the Pauli principle. The total spin-isospin Young diagrams are the result of direct internal product $\{f\}_{ST} = \{f\}_S \otimes \{f\}_T$ [34,35]. The corresponding classification for the $p^7$Be channel has been carried out in [6,36], and is given in Table 1.

Table 1. Classification of the orbital states for $N^7$Li ($N^7$Be) systems with isospin $T = 0$ and 1. Here $T$, $S$ and $L$ are the isospin, spin and orbital momenta of a two-particle state, respectively; $\{f\}_S$, $\{f\}_T$, $\{f\}_{ST}$, and $\{f\}_L$ are the spin, isospin, spin-isospin, and allowed orbital diagram $\{f_2\}_L$ [35]; and $\{f\}_{AS}$, $\{f\}_{FS}$ are the Young diagrams of the allowed and forbidden states. The conjugated Young diagrams are marked in **bold font**.

| Initial state $\{f_1\}_L+\{f_2\}_L$ | $T$ | $S$ | $\{f\}_T$ | $\{f\}_S$ | $\{f\}_{ST}$ | $\{f\}_L$ | $L$ | $\{f\}_{AS}$ | $\{f\}_{FS}$ |
|---|---|---|---|---|---|---|---|---|---|
| $p^7$Li $n^7$Be {1}+{43} | 0 | 1 | {44} | {53} | {71}+{611}+{53}+{521}+{431}+ +{4211}+{332}+**{3221}** | {53} {44} **{431}** | 1,3 0,2,4 1,2,3 | – – {431} | {53} {44} – |
| | | 2 | {44} | {62} | {62}+{521}+{44}+{431}+{422}+ +{3311} | {53} {44} {431} | 1,3 0,2,4 1,2,3 | – – – | {53} {44} {431} |
| $p^7$Li $n^7$Be $p^7$Be $n^7$Li {1}+{43} | 1 | 1 | {53} | {53} | {8}+2{62}+{71}+{611}+{53}+2{521}+ {5111}+ {44}+{332}+ 2{431}+ 2{422}+ {4211}+{3311}+**{3221}** | {53} {44} **{431}** | 1,3 0,2,4 1,2,3 | – – {431} | {53} {44} – |
| | | 2 | {53} | {62} | {71}+{62}+{611}+2{53}+2{521}+ +2{431}+{422}+{4211}+{332} | {53} {44} {431} | 1,3 0,2,4 1,2,3 | – – – | {53} {44} {431} |

The total Young diagram of a certain nucleus is defined in a similar way to the spin-isospin symmetry as the direct internal product of the orbital and spin-isospin parts $\{f\} = \{f\}_L \otimes \{f\}_{ST}$ [3]. The total wave function of the system does not lead to zero under antisymmetrization if it contains the antisymmetric component $\{1^A\}$, which is realized when the conjugate diagrams $\{f\}_L$ and $\{f\}_{ST}$ are multiplied. The orbital diagram $\{f\}_L$ conjugated to the spin-isospin diagram $\{f\}_{ST}$ is therefore considered to be allowed in this channel, and all other possible orbital symmetries are forbidden, since they lead to the zero wave function of the system of nuclear particles after its antisymmetrization [3,34–37].

In the present case, when the orbital diagram $\{43\}$ is assumed for $^7$Be, the $p^7$Be system with spin $S = 1$ contains forbidden levels with the $\{53\}$ diagram in $^3P$-waves, and the $\{44\}$ diagram in the $^3S_1$ wave.

The allowed $^3P$ state has the $\{431\}$ diagram, as shown in Table 1. Thus, $p^7$Be potentials should have a forbidden $\{44\}$ state in the $S_1$ wave and forbidden and allowed bound levels in $^3P$ waves, with Young diagrams $\{53\}$ and $\{431\}$, respectively. The ground state of $^8$B corresponds to an allowed $^3P_2$ configuration with a FS and a binding energy of −0.1375 MeV [6].

For spin $S = 2$, the allowed levels are generally absent, and all of the Young diagrams listed above correspond to the FS (see Table 1). Hence, we consider the GS of $^8$B in the $p^7$Be channel as a pure $^3P_2$ state, i.e. without a $^5P_2$ wave impurity. In this case, electromagnetic transitions are possible only between states with spin $S = 1$.

## 3. Potential description of elastic scattering phase shifts

We were not able to find data on the phase shifts of $n^7$Li, $n^7$Be or $p^7$Be elastic scattering at astrophysical energies [38], and therefore the scattering potentials in the $p^7$Be channel have been constructed here by analogy with $p^7$Li scattering [6]. We recall that the following moments $J^\pi, T = 3/2^-, 1/2$ [39] are known for the ground state of $^7$Be. Let us consider the structure of the resonance states of $^8$B [33], which is regarded as a cluster $p^7$Be system:

1. The first excited level at an excitation energy of 0.7695(2.5) keV or 632(±2.5) keV in c.m. corresponds to the first resonance state of $^8$B with $J^\pi, T = 1^+, 1$ and a width of 35.6(6) keV in c.m. on the spectrum of $^8$B levels (see Fig. 3 and Table 8.15 in [33]). This level can be compared to the $^3P_1$ state of the $p^7$Be system. Thus, the $^3P_1$ phase shift of the elastic $p^7$Be scattering should reveal resonance behavior at an energy of 630(3) keV in c.m. with a width of 35.7(6) keV (see Table 8.17 in [33]). This resonance can be seen in the dipole magnetic $M1$ transition $^3P_1 \to ^3P_2(GS)$.

2. The second resonance at an excitation energy of 2.32(2) MeV or 2.18(2) MeV in c.m. corresponds to the quantum numbers $J^\pi, T = 3^+, 1$ [33] and a width of 350(30) keV (cm). It can be matched to the $^3F_3$ state of the $p^7$Be scattering and the $^3F_3$ phase shift has the resonance form. This is revealed in the $E2$ multipole transition $^3F_3 \to ^3P_2(GS)$. We consider this transition in order to estimate the value of its cross section.

3. The third resonance level of $J^\pi, T = 2^-$ [33] with excitation energy of 3.5(5) MeV, which can be matched to the $^3D_2$ wave, has a width of 8(4) MeV (c.m.). With respect to the channel threshold, it has an energy of 3.36 MeV in c.m. This resonance can be seen in the electric dipole $E1$ transition $^3D_2 \to ^3P_2(GS)$.

4. The fourth resonance level in the $^8$B spectrum [33] corresponds to an excitation energy of 10.619(9) MeV with $J^\pi, T = 0^+, 2$ and a width smaller than 60 keV in c.m. (as shown in Fig. 3 and Table 8.15 of [33]). This resonance is not taken into account here, since we consider only an energy range of up to 5 MeV in c.m.

Note that in a more recent paper [40], several other additional levels of $^8$B are given and the energies of some others are refined, although the characteristics of the first resonance have not changed. We give a description of these in more detail, as follows:

1. After the first resonance, a new level with $J^\pi = 0^+$ is identified at an excitation energy of 1.9(1) MeV with a width of 530(60) keV, which can be matched to the $^3P_0$ state. This can lead to the $E2$ transition to the GS. We do not consider this in the present calculations due to the smallness of the partial cross sections of $E2$ processes.

2. The characteristics of the second resonance at 2.32 MeV with $J^\pi, T = 3^+, 1$, corresponding to the $^3F_3$ state, are somewhat refined. The value of 2.31(2) MeV is given at a width of 340(3) keV for the excitation energy, which differs little from the data in [33].

3. The next new level with $J^\pi, T = 2^+$ is at an excitation energy of 2.55(2) MeV or 2.41(2) MeV in c.m. with a width of 360(20) keV, which can be matched to the $^3P_2$ state. It can lead to the $M1$ transition $^3P_2 \to {}^3P_2(GS)$. Although no resonances are observed in this energy region in the experimental cross sections, the interaction potential for this state is constructed and its effect on the total cross sections of the $p^7$Be capture is considered.

4. There is another resonance with $J^\pi, T = 1^+$ at an excitation energy of 3.3(2) MeV or 3.16(20) MeV in c.m. with a width of 1.34 MeV, which can be matched to the $^3P_1$ state. This can lead to the $M1$ transition $^3P_1 \to {}^3P_2(GS)$. Since no resonance is observed in this energy range in the experimental cross sections, we do not consider its effect here, although we give the corresponding interaction potential, which may be appropriate for future refinements.

5. The resonance described above at 3.5(5) MeV with $J^\pi, T = 2^-$ and width 8(4) MeV is now located at a higher excitation energy of 3.8 MeV or 3.66 MeV in c.m. with a width almost half that at 4.7 MeV. To construct this potential, these exact parameters are used for the resonance.

In [40], several levels are given at excitation energies of above 5 MeV, but their momenta are not precisely determined, and preventing them from being matched to a definite partial wave; hence, we do not consider them, since it is impossible to construct the relevant intercluster potentials in this case.

All partial intercluster potentials are chosen in a simple Gaussian form with a point-like Coulomb term

$$V(r, JLS\{f\}) = -V_0(JLS\{f\})\exp\{-\alpha(JLS\{f\})r^2\}. \qquad (7)$$

Here, $V_0$ and $\alpha$ are the parameters of the potential, and are determined on the basis of the description of the elastic scattering or resonance phase shifts in the spectra of a finite nucleus in certain partial waves for scattering processes or the characteristics of the bound state of the $p^7$Be system for $^8$B. In the present approach, all potentials are considered to be dependent on the quantum numbers $JLS$ {$f$}; that is, the potentials depend on those characteristics on which the scattering phase shifts or the excited and resonance levels in the final nucleus depend. In other words, for each partial wave corresponding to the states with different Young diagrams {$f$}, the interaction potentials may have different parameter values [4,5,25]. This is one of the modifications of the MPCM used here, and means that all potentials explicitly depend on Young diagrams.

By analogy with $p^7$Li scattering and based on the data in [33], we assume that the $^3S_1$ scattering phase shift in the region up to 5 MeV is practically equal to zero, since there are no resonance levels with negative parity with $J = 1$ in the spectrum of $^8$B. Thus, a near-zero phase shift is obtained with the Gaussian potential and parameters given above and listed in the first row of Table 2.

Table 2. Potential parameters of the $p^7$Be interaction

| No. | $^{(2S+1)}L_J$ | $V_0$, MeV | $\alpha$, fm$^{-2}$ | Resonance energy, $E_{cm}$, MeV | Resonance width, $\Gamma_{cm}$, keV | $C_w$ | Radius $\langle r^2 \rangle^{1/2}$, fm |
|---|---|---|---|---|---|---|---|
| 1 | $^3S_1$ | 343.0 | 1.0 | --- | --- | --- | --- |
| 2 | $^3P_1$ | 709.85 | 0.83 | 0.632 | 35.9 | --- | --- |
| 3 | $^3P_2$ | 3346.915 | 4.0 | 2.41 | 360 | --- | --- |
| 4 | $^3P_0$ | 580.0 | 1.0 | --- | --- | --- | --- |
| 5 | $^3F_3$ | 104.555 | 0.055 | 2.18 | 351 | --- | --- |
| 6 | $^3D_2$ | 116.04 | 0.095 | 3.66 | 4700 | --- | --- |
| 7 | $^3P_2$ | 330.414634 | 0.375 | --- | --- | 1.76(1) | 2.84 |
| 8 | $^3P_2$ | 70.484737 | 0.21 | --- | --- | 1.77(1) | 2.83 |
| 9 | $^3P_1$ | 146.0658 | 0.47 | 0.632 | 35.9 | --- | --- |
| 10 | $^3P_2$ | 676.16 | 2.3 | 2.41 | 358 | --- | --- |
| 11 | $^3F_3$ | 39.364 | 0.04 | 2.18 | 352 | | |
| 12 | $^3D_2$ | 29.95 | 0.062 | 3.66 | 4600 | --- | --- |
| 13 | $^3S_1$ | 300.0 | 1.0 | --- | --- | --- | --- |

This potential contains a bound FS for the {44} diagram, as follows from the classification of states in Table 1, and this leads to a scattering phase shift of ±1.0° in the energy region up to 5 MeV (c.m.). It is shown in Fig.1a by a black solid curve.

Furthermore, for the non-resonance $^3D_1$ and $^3D_3$ scattering waves, a potential was used at $L = 2$ with parameters labeled 1 in Table 2, which also has a FS for the {44} diagram. Here, we give the phase shift from zero degrees, i.e. in a more familiar form; however, we must remember that since it has a FS, then in accordance with the generalized Levinson theorem [3] this phase shift should start from 180°.

A direct test, i.e. the use of alternative potentials, including those with zero-depth parameters, shows that such changes do not in practice affect the shape and magnitude of the astrophysical $S$-factor of the radiative capture.

The most important aspect is that the $^3S_1$ phase shift at low energies is very close to zero. If we ignore the classification according to Young diagrams and the presence of a FS, the potential of the $^3S_1$ wave without the FS will simply have zero depth. In this case, in calculations of the *S*-factor, we account for the non-resonance $^3D_1$ and $^3D_3$ scattering waves corresponding to the zero potential depth at $L = 2$, which no longer has an FS.

In the $^3P_1$ scattering wave, there is an above-threshold level with an excitation energy of 0.7695 MeV and $J^\pi, T=1^+, 1$ [33]. The 0.7695 MeV level has a width of 35.7(6) keV (c.m.) for the capture reaction $p^7\text{Be} \to {}^8\text{B}\gamma$ and elastic $p^7\text{Be}$ scattering (see Table 8.17 in [33]), which is three times larger than the analogous width of the resonance in the $p^7\text{Li}$ system. This resonance results in a smoother increase in the $^3P_1$ phase shift of the $p^7\text{Be}$ elastic scattering than in the $p^7\text{Li}$ system [6], and the resonance scattering phase shifts can be described by a Gaussian potential (labeled No. 2 in Table 2). This potential has a bound FS for the {53} diagram, and the calculation results for the $^3P_1$ scattering phase shift are shown in Fig. 1a by a green solid curve.

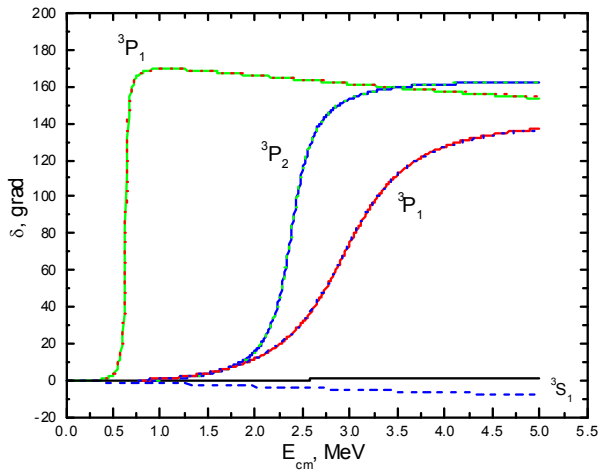 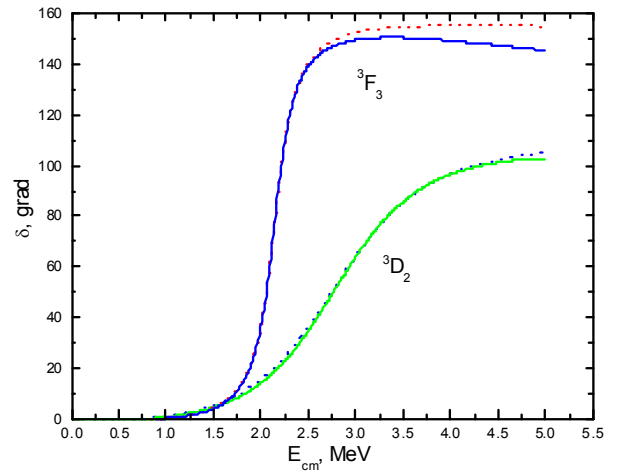

Fig.1a. $^3S_1$ and $^3P_J$ phase shifts of the $p^7\text{Be}$ elastic scattering

Fig.1b. $^3F_3$ and $^3D_2$ phase shifts of the $p^7\text{Be}$ elastic scattering

We can use the parameters labeled No. 3 in Table 2 for the potential of the resonance $^3P_2$ wave. This potential has a bound FS, and the calculation results of the $^3P_2$ scattering phase shift are shown in Fig. 1a by the blue solid curve.

For the potential of the second resonance specified in the $^3P_1$ wave, we suggest the following set of parameters:

$$V_0 = 1038.985 \text{ MeV and } \alpha = 1.25 \text{ fm}^{-2}. \tag{8}$$

The potential has a bound FS, the resonance energy is 3.16 MeV at a width of 1.34 MeV (c.m.), and the results of the calculated $^3P_1$ scattering phase shift are shown in Fig. 1a by the red solid curve.

For the potential of the non-resonating $^3P_0$ wave with a FS (which is not accounted for in further calculations since it leads to the *E*2 transition to the GS and is assumed to be small in the treating energy region), it is possible to use the parameters labeled No. 4 in Table 2, which lead to scattering phase shifts of $\pm 0.6°$ in the region up to 5 MeV (c.m.). For comparison, we can use the potential of the $^3P_0$ wave without FS, which simply has zero depth.

For the resonance $^3F_3$ phase shift the potential accounting the FS for the {53} diagram and with parameters No. 5 from Table 2 was found, and its phase shift is shown in Fig. 1b by the blue solid curve.

For the resonance $^3D_2$ scattering phase shift, it is possible to obtain the potential using the FS for the {44} diagram and the parameters labeled No. 6 in Table 2, and its phase shift is shown in Fig. 1b by the green solid curve.

For the potential of the bound $^3P_2$ state of the $p^7$Be system, which refers to the ground state of $^8$B in the cluster channel under consideration, the parameters labeled No. 7 are used. For this potential, a binding energy of −0.137500 MeV [33] with an accuracy of $10^{-6}$ MeV [41] was obtained. The AC error given in Table 2 is determined by averaging over the interval 5–25 fm, where the asymptotic constant remains relatively stable. In addition to the allowed bound state (BS) corresponding to the GS of $^8$B, this $^3P_2$ potential has an FS in accordance with the classification of the orbital states given in Table 1.

The *rms* radius of $^8$B is expected to be somewhat larger than the $^9$Be radius, which is equal to 2.519(12) fm [33], since $^8$B is very weakly bound in the $p^7$Be channel (with a binding energy of −0.1375 MeV). Thus, the *rms* of this "loose" state can be assumed to be close to 3 fm [33]. The obtained value of 2.84 fm for the *rms* radius of $^8$B in the GS is therefore a reasonable value for the $p^7$Be cluster configuration.

For the dimensional asymptotic constant of the $p^7$Be channel, the result is 0.613(82) fm$^{-1}$ or 0.783(51) fm$^{-1/2}$ [19]. Dimensional and dimensionless constants are defined above, assuming the spectroscopic $S_f$ factor to be equal to unity [15]. Then, at a value of 0.391 for the GS of $^8$B in the $p^7$Be channel, the dimensionless AC is equal to 2.00(14). For AC, the value 0.57(10) fm$^{-1}$ is given in [18], i.e. 0.755(69) fm$^{-1/2}$, which gives 1.93(18) after recalculation. The potential labeled as No. 7 in Table 2 gives AC, which is at the lower limit of this quantity. In addition, a theoretical value of about 1.1 is given for the spectroscopic factor $S_f$ in [21], and a value of 0.915 is obtained in [42]. On average, $S_f$ can be taken as 1.0(1), which expands the range of possible values of AC.

Since the asymptotic normalization coefficient ANC is equal to $A_{NC} = \sqrt{S_f} \cdot C$, then on the basis of the above values, we can use a range of values from 0.686 to 0.834 fm$^{-1/2}$ with an average value of 0.760(74) fm$^{-1/2}$. In this case, we have a range of 0.653 to 0.869 fm$^{-1/2}$ for the dimension AC, with an average value of 0.76(11) fm$^{-1/2}$. We then obtain the range of values 1.67–2.22 for the dimensionless $C_W$, and the value of the potential AC labeled No. 7 is completely consistent with this.

Rather than using the above classification based on Young diagrams to compare the results obtained below, we can use the potential of the bound ground $^3P_2$ state of the $p^7$Be system without FSs, labeled No. 8 in Table 2. This potential gives a binding energy of −0.137500 MeV with an accuracy of $10^{-6}$ MeV, and the *rms* charge radius and a dimensionless asymptotic constant are listed in Table 2.

The potential of the resonance $^3P_1$ wave without FSs has the parameters labeled No. 9 in Table 2. This potential leads to the $^3P_1$ scattering phase shift shown by the red dotted curve in Fig. 1a.

The parameters labeled No. 10 in Table 2 were found for the potential of the resonance $^3P_2$ wave without FSs. This potential leads to the $^3P_2$ scattering phase shift shown by the green dotted curve in Fig. 1a.

For the potential of the second resonance in the $^3P_1$ wave, the corresponding parameters are determined as follows:

$$V_0 = 201.05 \text{ MeV and } \alpha = 0.7 \text{ fm}^{-2}. \qquad (9)$$

This potential leads to a resonance energy of 3.16 MeV at a width of 1.35 MeV (c.m.), and the calculation results of the $^3P_1$ scattering phase shift are shown by the blue dotted curve in Fig. 1a.

For the resonance $^3F_3$ phase shift, it is possible to obtain a potential with the parameters labeled No. 11 in Table 2. This leads to the phase shift shown by the red dotted curve in Fig 1b.

For the resonance $^3D_2$ scattering phase shift, it is possible to obtain a potential without an FS, with the parameters labeled No. 12. This leads to the phase shift shown by the blue dotted curve in Fig. 1b.

It should be noted that the $^3S_1$ scattering phase shift is obtained in [20], which drops below zero at 3 MeV and has a value of about −5°. This phase shift is reproduced by the parameters labeled No. 13 in Table 2. This potential will be used later to compare the results, and its phase shift is shown by the blue dashed curve in Fig. 1a. In this version of the calculations, this potential was used for $L = 2$ for non-resonance $^3D_1$ and $^3D_3$ scattering waves.

## 4. Astrophysical S-factor of the $^7$Be$(p,\gamma)^8$B capture reaction

Some of the most recent experimental measurements of the $S$-factor of the $^7$Be$(p,\gamma)^8$B reaction in the energy range 100 keV to 2.5 MeV, which are in good agreement with each other, were performed in [17,43,44]. In [17], the data reported by the authors of the earlier paper are refined. Once again, we note that dozens of measurements of the total cross sections of this reaction have been performed [45], starting in the 1960s; however, the results of different measurements can differ by up to 30–40%, as demonstrated in [20,21].

In considering the electromagnetic transitions in the $^7$Be$(p,\gamma)^8$B reaction, we are taking into account the $E1$ capture from the $^3S_1$ scattering wave to the ground bound $^3P_2$ state of $^8$B in the cluster $p^7$Be channel. Other transitions considered are listed in Table 3. The $E2$ transitions from the $^3P$ waves of the $p^7$Be scattering to the bound $^3P_2$ state of $^8$B are also considered; it was found that these were two to three orders of magnitude smaller than for the $E1$ transition from the $^3S_1$ wave.

Table 3. Classification of considered transitions to the $^3P_2$ GS, as defined by potentials No. 7 and No. 8, algebraic coefficients $P_J^2$ in the corresponding total cross sections, excitation energy values, and relevant continuum potential sets

| No. | [$^{(2S+1)}L_J$]$_i$ | $NJ$ | $P^2$ | Excitation energy, MeV | Continuum potentials |
|---|---|---|---|---|---|
| 1 | $^3S_1$ | $E1$ | 5 | – | No.1 and zero depth; No.13 |
| 2 | $^3P_1$ | $M1$ | 5/2 | 0.7695 [33] | No.2 and No.9 |
| 3 | $^3P_2$ | $M1$ | 15/2 | 2.55 [40] | No.3 and No.10 |
| 4 | $^3D_1$ | $E1$ | 1/10 | – | No.1 and zero depth; No.13 |
| 5 | $^3D_2$ | $E1$ | 3/2 | 3.8 [40] | No.6 and No.12 |
| 6 | $^3D_3$ | $E1$ | 42/5 | – | No.1 and zero depth; No.13 |
| 7 | $^3F_3$ | $E2$ | 9/5 | 2.32 [33] | No.5 and No.11 |

To calculate the astrophysical $S$-factor, we use the standard expression [46,47]

$$S(NJ, J_f) = \sigma(NJ, J_f) E_{cm} \exp\left(\frac{31.335 Z_1 Z_2 \sqrt{\mu}}{\sqrt{E_{cm}}}\right), \quad (10)$$

where $\sigma$ is the total cross section of the radiative capture process in barn, $E_{cm}$ is the relative particle energy in keV (c.m.), and the other symbols are defined above. A numerical coefficient of 31.335 is obtained on the basis of the modern values of the fundamental constants [48].

The results for the calculated astrophysical $S$-factor using the potentials given above are shown in Fig. 2a, with the potential labeled No. 7 in Table 2 used for the GS. The black dotted curve shows the $E1$ transition labeled No.1 in Table 3, and the red solid curve is the sum of all $E1$ and $M1$ transitions (Nos. 1–7), which closely match the experimental data shown in Fig. 2a. As potentials of the continuous spectrum, we use the sets of parameters leading to FS numbered in Table 2.

At an energy of 10 keV, the $S$-factor was found to be 20.7(1.0) eV·b. The error arises from the accuracy of the calculations, and is estimated at 3–5%. The blue dashed curves show transitions Nos. 4–6 from Table 3 for the $D$ scattering waves. The red dotted curves show cross sections for the transitions from $P$ waves Nos. 2 and 3, where the scattering $^3P_2$ wave is assumed to be non-resonance and is fitted by parameters labeled No. 4 in Table 2. At the very bottom of Fig. 2a, the green dashed curve shows the results for the $E2$ transition (No. 7 in Table 3), which makes practically no contribution to the total capture cross sections.

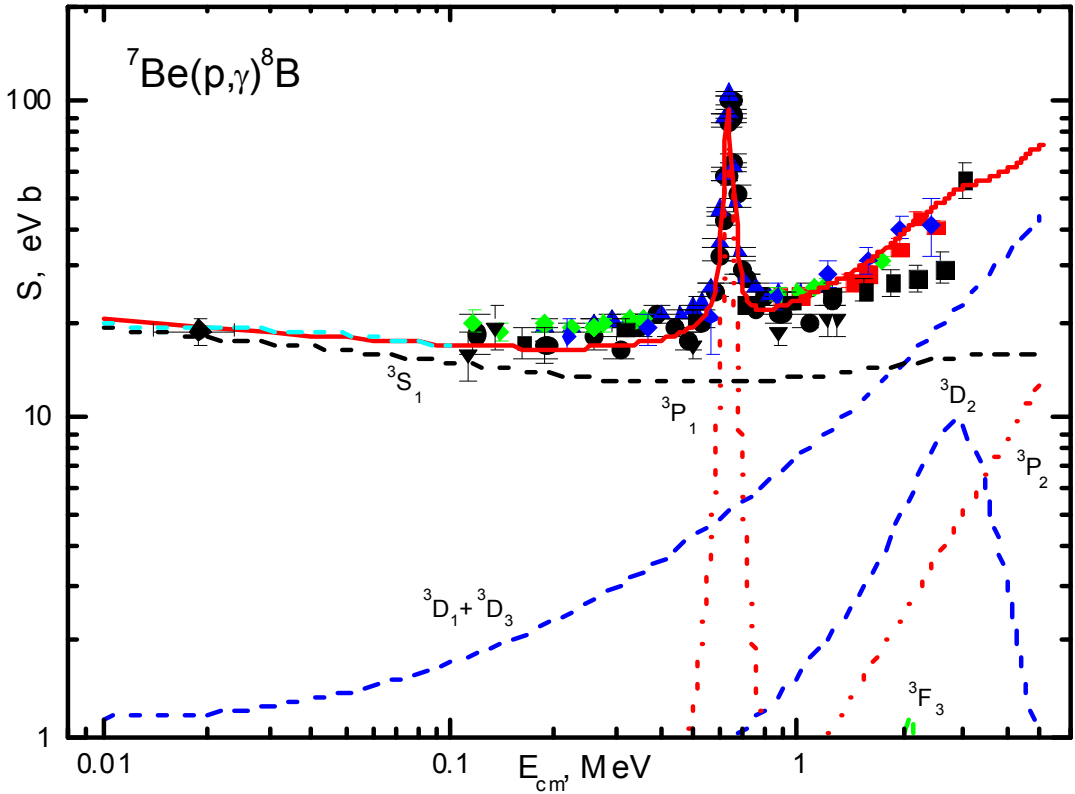

Fig. 2a. Astrophysical $S$-factor for radiative proton capture by $^7$Be. The points are experimental data from [43], blue triangles, red squares and green rhombuses from [17], black squares from [14], black triangles from [49], blue rhombi from [50], black rhombi at 19 keV from [51]. The curves show the calculation results for different electromagnetic transitions with the potentials given in the text.

It can be seen from Fig. 2a that the resonance in the *S*-factor reaches 95 eV·b at an experimental value of 105.5(2.2) eV·b [17]. In the energy range below resonance, the results calculated here are in complete agreement with the experimental data presented in [17,43]. The main contribution is given by the process labeled No. 1 in Table 3. Above the first resonance due to process No. 2, the main contribution to the correct description of the experimental measurements is given by processes No. 1 and Nos. 4–6 from Table 3. Using the $^3S_1$ potential No. 13 rather than No. 1 from Table 2 does not lead to changes in the total capture cross sections shown in Fig. 2a.

As the present paper was being prepared, new experimental data appeared in [51], in which the astrophysical *S*-factor was extracted from the solar neutrino flux at lowest, from earlier measured, energy 19 keV and a value of 19(1.8) eV·b was assumed for the *S*-factor. The results of these measurements are given in Fig. 2 by the black rhombi. As can be seen, these are within the limits of error and show good agreement with our calculations at this energy. In particular, a value of 19.4 keV·b was obtained in our calculations at an energy of 20 keV, as shown in Fig. 2a.

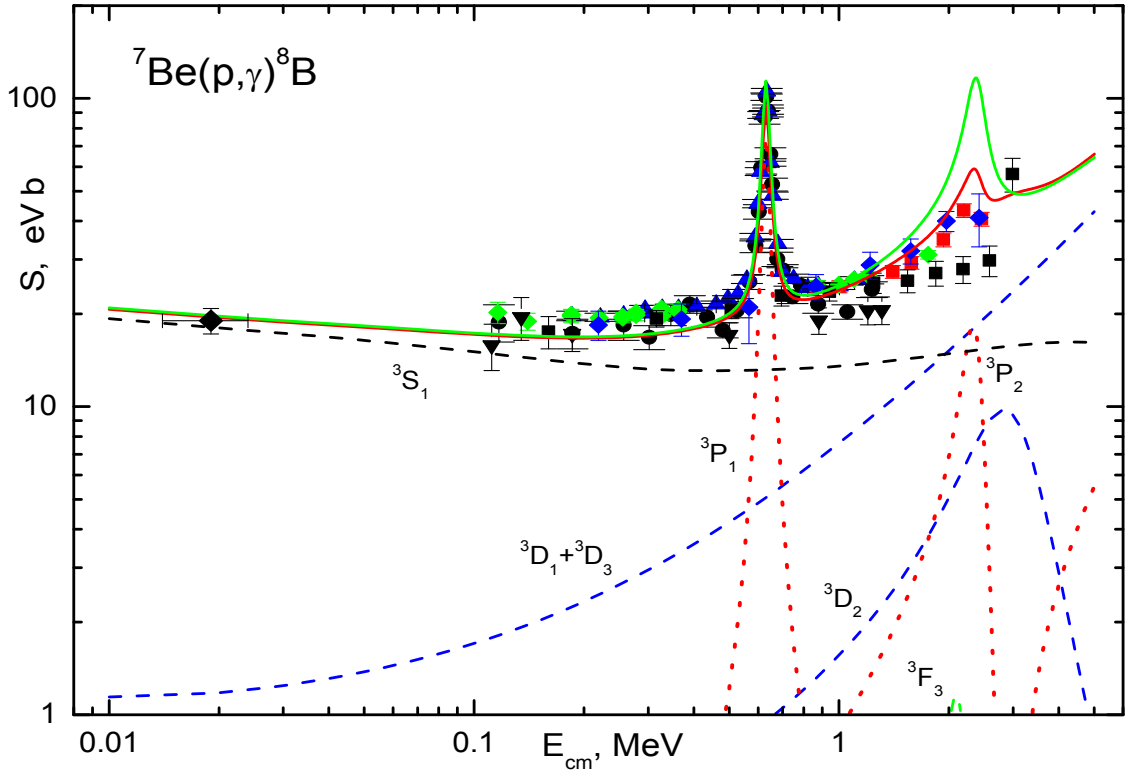

Fig. 2b. Astrophysical *S*-factor for proton capture by $^7$Be (notation as in Fig. 2a)

The same transitions were also considered for potentials without bound FSs. As the GS and resonating continuous waves, we use the sets of parameters Nos. 8, 9, 11 and 12 from Table 2. Zero depth potentials are used for the non-resonance $^3S_1$ and $^3P_2$ waves. As a result, the *S*-factor was found to be 20.9(1.0) eV·b at an energy of 10 keV. The resonance of the *S*-factor has a value of 114 eV·b, and the general shape of the calculated *S*-factor is virtually the same as the results shown in Fig. 2a. After the first resonance (i.e. at higher energies) the results of these two types of potential are almost the same. The calculation results for the *S*-factor and their comparison with experimental results do not give an unambiguous answer to the question of the presence or absence of an FS in the potentials of the $p^7$Be interaction. Both types of interaction lead to similar results, and the difference is therefore observed only in the magnitude of the peak of the first resonance.

In an energy range of 10 to 100 keV, the calculated *S*-factor shown in Fig. 2a can be approximated by a simple expression:

$$S(E) = S_0 + S_1 E$$

with parameters $S_0 = 20.1$ and $S_1 = -33.0$. Here, the energy *E* is given in MeV. The results of this parameterization with $\chi^2 = 0.16$ with errors of 5% in the theoretical calculations are shown in Fig. 2a by the light blue dashed curve.

If we use the resonance potential labeled No. 3 in Table 2 with an FS, then the calculated cross sections show a slight increase in the energy range of 2.4 MeV. The results are shown by the red solid curve in Fig. 2b. The resonance character of the $^3P_2$ capture cross section can clearly be seen. For potentials with FSs, its magnitude reaches 59 eV·b.

Note that when using potentials without FS and interaction No. 10 from Table 2, this resonance for the $^3P_2$ wave reaches 117 eV·b and is clearly inconsistent with the available experimental data. The results for the total cross section are shown in Fig. 2b by the green solid curve. Hence, it can be seen that potentials with an FS are preferable, since they more accurately convey the behavior of the experimental cross sections at energies above the first resonance.

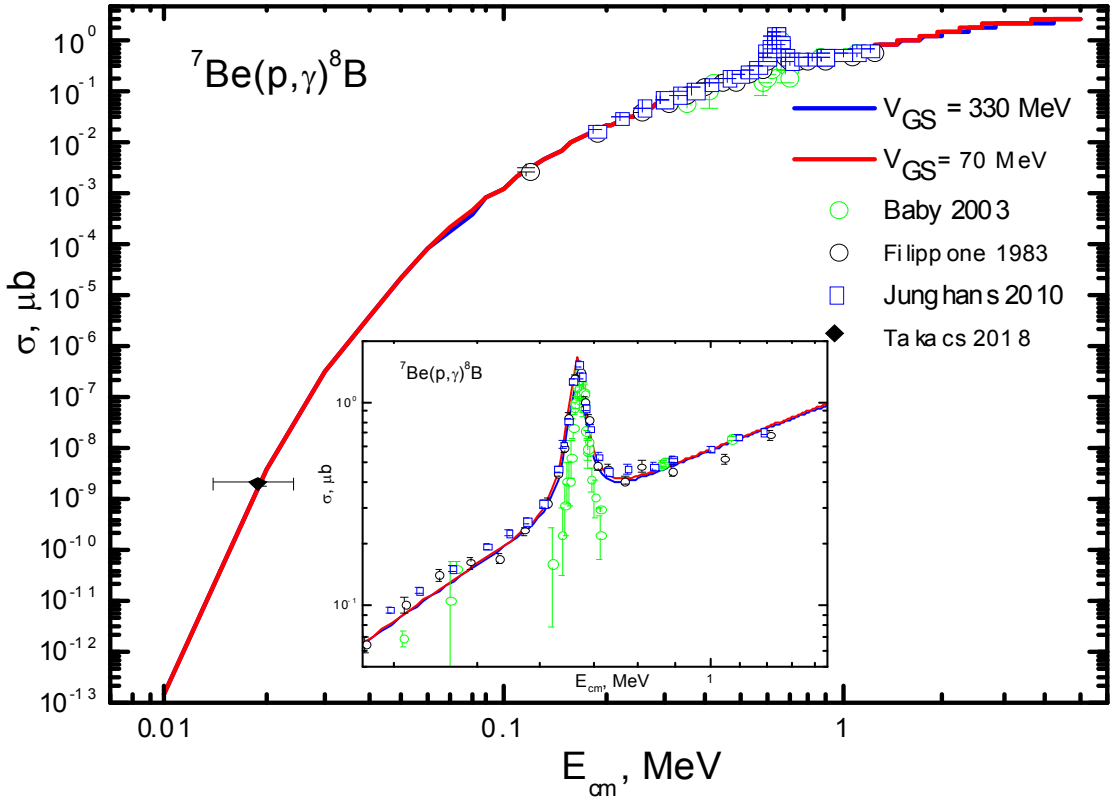

Fig 3. Total cross sections for the proton capture on $^7$Be with two options of GS potentials.

Fig. 3 illustrates the total cross section calculated for the proton capture by $^7$Be at energies of up to 5 MeV in c.m. We use this cross section to calculate the reaction rate, which is shown in Fig. 4 at temperatures from 0.01 to 10.0 $T_9$. Both figures show the results for the deep and shallow sets of interaction potentials, which correspond to the *S*-factor given in Fig. 2a. Fig. 3 clearly shows the resonance at 0.632 MeV in c.m., which is due to the resonance of the $^3P_1$ scattering wave. The experimental data in Fig. 3 are taken from [17,43,51,52].

## 5 Reaction rate

The reaction rate is defined in the standard way [29]:

$$N_A \langle \sigma v \rangle = 3.7313 \cdot 10^4 \mu^{-1/2} T_9^{-3/2} \int_0^\infty \sigma(E) E \exp(-11.605 E / T_9) dE ,$$

where *E* is in MeV, the cross section σ(*E*) is in μb ($10^{-6}$·b), μ is the reduced mass in amu, and $T_9$ represents $10^9$ K.

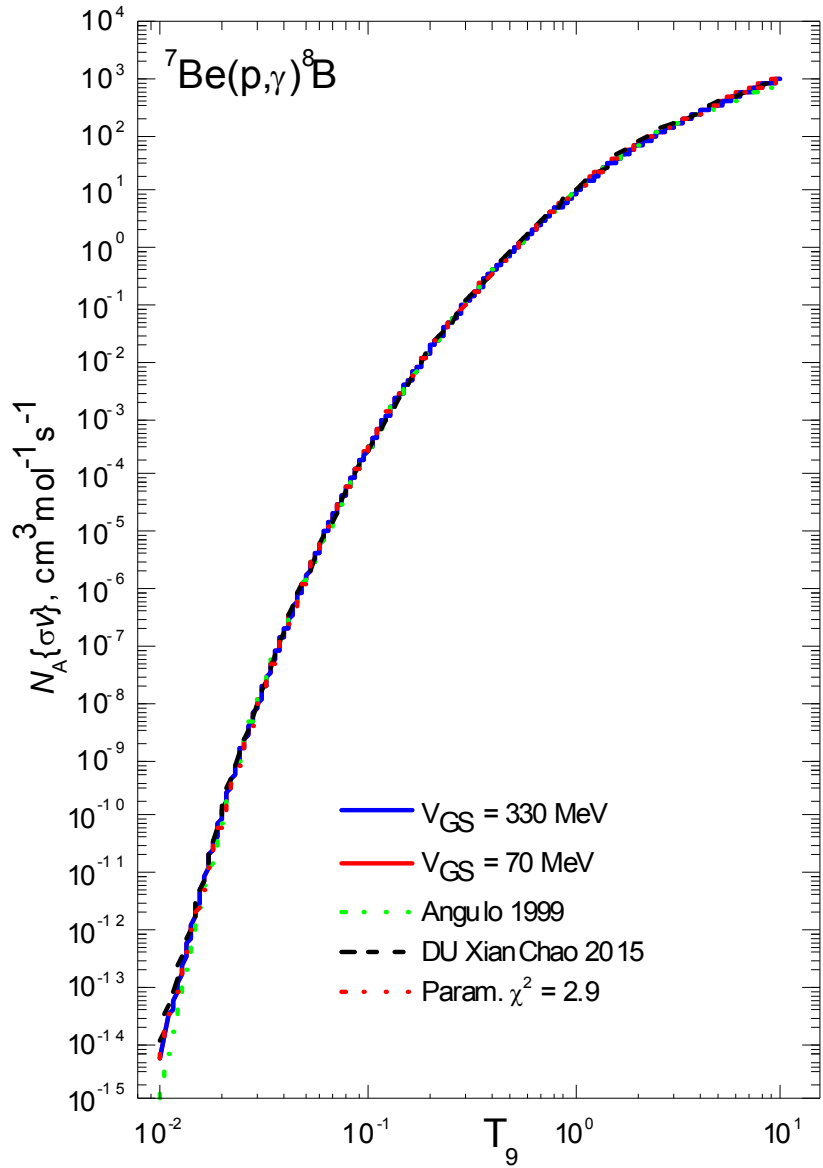

Fig. 4. Reaction rate for proton capture by $^7$Be (notation for the curves as in Fig. 2)

The green dotted curve in Fig. 4 shows the results of the calculated reaction rate from [29], which differs from our results only at the lowest temperatures. The black dashed curve in Fig. 4 is taken from [18], and almost coincides with the present results.

This form of the reaction rate, which is used for potentials with an FS, for example, is parameterized by the expression

$$N_A \langle \sigma v \rangle = 81.15034 / T_9^{2/3} \cdot \exp(-10.3934 / T_9^{1/3}) \cdot (1.0 + 20259.6 \cdot T_9^{1/3} -$$
$$- 29831.13 \cdot T_9^{2/3} + 7725.04 \cdot T_9 + 3133.587 \cdot T_9^{4/3} - 1134.655 \cdot T_9^{5/3}) +$$
$$+ 73867.4 / T_9^{1/2} \exp(-9.01236 / T_9^{1/2})$$

with $T_9$ in units of $10^9$ K, giving $\chi^2 = 2.9$ with an 1% error of in the theoretical reaction rate. The results of this approximation are shown in Fig. 4 by the red dotted curve.

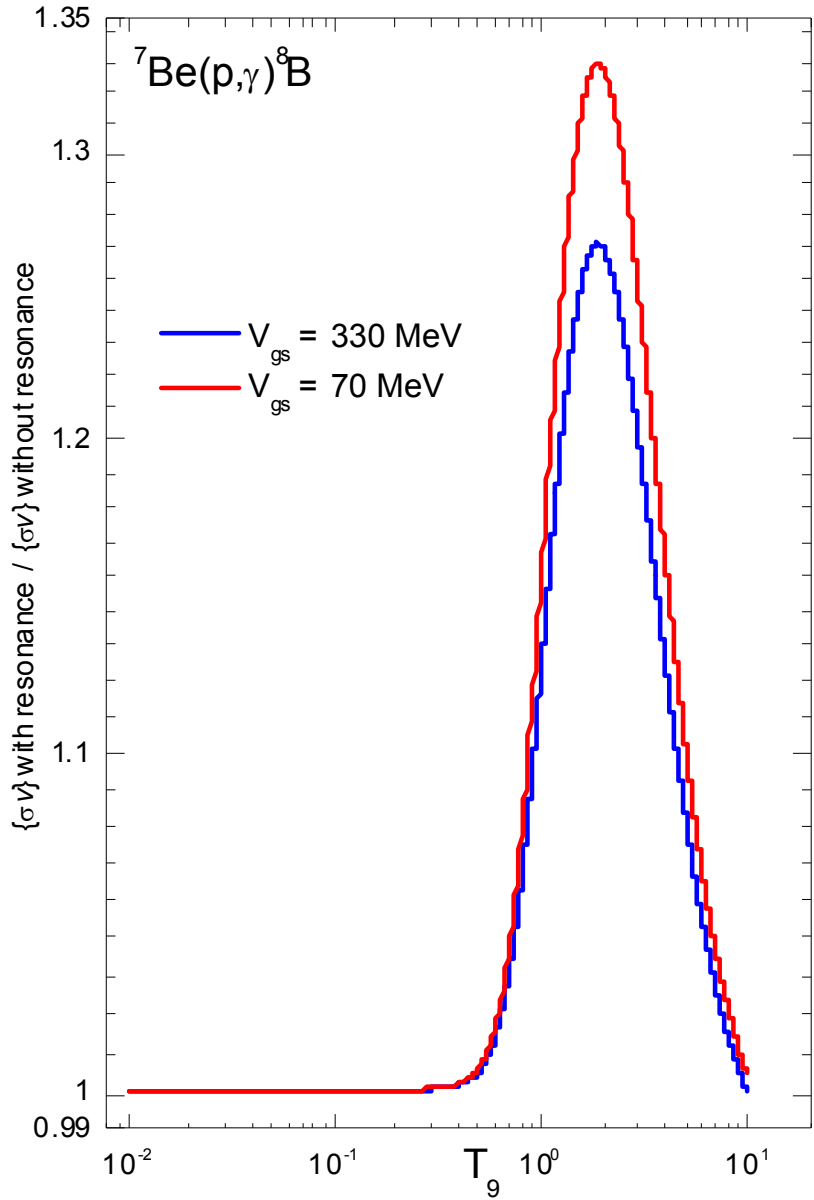

Fig. 5. The ratio of the reaction rate in the presence of a resonance at 0.632 MeV to the reaction rate without taking into account this resonance for deep and shallow potentials

Fig. 5 shows the ratio of the reaction rates of the proton capture by $^7$Be, with taking into account the resonance at 0.632 MeV and without its accounting. The blue curve shows the ratio for the GS and scattering potentials with FS.

It can clearly be seen from Fig. 5 that the contribution of the resonance at 0.632 MeV for a temperature of about 1.9 $T_9$ varies from about 27% to 34% for different types of potential, and has a significant effect on the shape of the reaction rate. This change in the reaction rate to greater values may have some influence on certain processes in the primordial nucleosynthesis of the Universe.

Within the range of solar temperatures (i.e. 0.01–0.02 $T_9$) the resonance part of the reaction rate exceeds the non-resonance part only by 0.015–0.02%. However, due to the Maxwellian tail of the energy distribution, the contribution of the resonance part to the $^8$B formation processes can be much larger. For example, at $T_9 = 1.0$, this contribution is already 12.5%, as shown by the blue curve in Fig. 5.

The importance of taking this resonance into account was demonstrated for the first time in [18], where the contributions of the reaction rates are given with and without this resonance. In particular, at 2.0 $T_9$, the contribution of the first resonance to the reaction rate was determined to be 34%. The current results fully confirm the numerical conclusions of the abovementioned work in terms of the importance of the contribution of this resonance to the rate of proton radiative capture reaction by $^7$Be.

## 6. Conclusions

As a result of an analysis performed on the basis of the results obtained here, the following conclusions can be drawn, which we have divided into two parts.

### 6.1. Physical aspects

In the modified potential cluster model [4,5,25], the *E*1, *E*2, and *M*1 transitions from various scattering waves to the $^3P_2$ ground state of $^8$B in the $p^7$Be channel are considered. For both types of potential, deep and shallow, it is possible to correctly describe the general behavior of the experimental data for the astrophysical *S*-factor at all energies considered.

A simple parameterization of the *S*-factor is given for the low-energy region. It was shown that, based on the model used, it is possible to closely describe the available experimental data in an energy range up to 3 MeV c.m.

Note, that in *ab initio* models, nodeless wave functions are used for the GS of the $p^7$Be system [20,21]. These studies do not consider the resonance at 0.632 MeV. At the same time, as can be seen from Fig. 2, a noticeable difference in the cross sections is observed only at the resonance energy; at energies above or below this resonance, the total cross sections for the two types of potentials considered here are almost exactly the same. In addition, the resonances for $n^7$Li and $p^7$Be capture were also excluded from the microscopic three-cluster model [53]. The first of these was considered by the current authors in [24], and the results for the second reaction are given in the present paper.

In spite of the similarity of the results for both types of potential, preference should be given to the FS potential, as it more accurately describes the *S*-factor at energies of 2−3 MeV, as shown in Fig. 2b. This potential completely agrees with the state classification given in Table 1.

Finally, it should be noted once again that the latest experimental data [51] became known to us only after finishing all our calculations. These data were simply plotted in Fig. 2, allowing us to demonstrate good agreement between these data and our calculations.

**6.2. Astrophysical aspects**

Based on the model used here, results for the reaction rate are obtained that are in good agreement with data from other, earlier studies. The parameterization of the calculated rate of the considered reaction is carried out in a simple form. A significant contribution of the resonance at 0.632 MeV (c.m.) to the reaction rate is demonstrated, which was first shown in [18]. Its influence is already felt at temperatures above 0.3–0.4 $T_9$ and continues to almost 10 $T_9$, reaching a maximum of about 30% at a temperature of about 1.9–2.0 $T_9$.

This significant influence of the resonance part of the reaction rate of $^7$Be(p,γ)$^8$B capture can lead to a certain increase in the number of $^8$B nuclei in the Sun. However, we have not yet had an opportunity to estimate the degree of increase in the yield of these nuclei. We therefore draw the attention of other research groups who have this opportunity to the results presented here. We hope that the results obtained here will help in refining input data for models of solar physics.

**Acknowledgments**

This work was supported by the Ministry of Education and Science of the Republic of Kazakhstan (Grant No. BR05236322 entitled "Study reactions of thermonuclear processes in extragalactic and galactic objects and their subsystems"), through the Fesenkov Astrophysical Institute of the National Center for Space Research and Technology of the Ministry of Defense and Aerospace Industry of the Republic of Kazakhstan (RK).